\documentclass[useAMS,usenatbib,usegraphicx]{mn2e}

\newcommand{\ergcms}{erg~cm$^{-2}$~s$^{-1}$}

\newcommand{\fek}{Fe~K$\alpha$}
\newcommand{\fel}{Fe~L$\alpha$}
\newcommand{\etal}{et al.}

\def\ltsima{$\; \buildrel < \over \sim \;$}
\def\simlt{\lower.5ex\hbox{\ltsima}}
\def\gtsima{$\; \buildrel > \over \sim \;$}
\def\simgt{\lower.5ex\hbox{\gtsima}}
\def\gsimeq
{\hbox{\raise0.5ex\hbox{$>\lower1.06ex\hbox{$\kern-1.07em{\sim}$}$}}}
\def\lsimeq
{\hbox{\raise0.5ex\hbox{$<\lower1.06ex\hbox{$\kern-1.07em{\sim}$}$}}}

\def\et{{et al.\ }}

\def\xmm{{\it XMM-Newton }}

\def\xmm{{\it XMM-Newton}}

\def\suzaku{{\it Suzaku}}

\def\iras{IRAS~13224--3809}

\def\pg1404{PG~1404+226}

\def\dchisq{\hbox{$\Delta\chi^2$}}

\def\mnras{MNRAS}
\def\apj{ApJ}
\def\apjs{ApJS}
\def\aap{A\&A}

\title[Relativistic reflection in \iras] 
 {Relativistic disc reflection in the extreme NLS1 \iras}

 \author[G.\ Ponti et al. ]
{G.~Ponti$^{1,2}$\thanks{ponti@iasfbo.inaf.it},
L. C. Gallo$^{3}$, A. C. Fabian$^{4}$, G. Miniutti$^{5}$, A. Zoghbi$^{4}$,
P. Uttley$^{2}$, 
\newauthor
R. R. Ross$^{6}$, R. V. Vasudevan$^{4,7}$, Y. Tanaka$^{8}$ and W. N. Brandt$^{7}$
\\ \\
   $^1$APC Universit\'e Paris 7 Denis Diderot, 75205 Paris Cedex 13, France \\
   $^2$School of Physics and Astronomy, University of Southampton, Highfield, 
   Southampton SO17 1BJ, UK\\
   $^3$Department of Astronomy \& Physics, Saint Mary's University, 923 
   Robie Street, Halifax, NS B3H 3C3, Canada \\
   $^4$Institute of Astronomy, Madingley Road, Cambridge CB3 0HA\\
   $^5$Centro de Astrobiologia (CSIC-INTA); LAEFF, P.O. Box 78, E-28691, 
   Villanueva de la Ca\~nada, Madrid, Spain\\
   $^6$Physics Department, College of Holy Cross, Worcester, MA 01610, USA \\
   $^7$Department of Astronomy and Astrophysics, The Pennsylvania State University, 
   525 Davey Lab., Univeristy Park, PA 16802, USA\\
   $^8$Max-Planck-Institut fur extraterrestrische Physik, Postfach 1603, D-85748 
   Garching, Germany\\
   }

\date{}

\pagerange{\pageref{firstpage}--\pageref{lastpage}}
\pubyear{2001}

\usepackage{times}

\begin{document}

\label{firstpage}

 \maketitle

\begin{abstract}
We present a spectral variability study of the \xmm\ and
\suzaku\ observations of one of the most extreme Narrow Line Seyfert 1
galaxies, \iras.  The X-ray spectrum is characterized by two main
peculiar features, i) a strong soft excess with a steep rise below
about 1.3 keV and ii) a deep drop in flux above 8.2 keV. 
  Although absorption--based interpretations may be able to explain
  these features by a suitable combination of ionization,
  covering factors, column densities and outflowing velocities, we
  focus here on a reflection--based interpretation which interprets
  both features, as well as the large soft excess, in terms of
  partially ionized reflection off the inner accretion disc. We show
  that the two peculiar spectral features mentioned above can be
  reproduced by two relativistic emission lines due to Fe K and Fe L.
The lines are produced in the inner accretion disc and independently
yield consistent disc parameters. 
We argue that the high L/K intensity ratio is broadly consistent with 
  expectations from an ionized accretion disc reflection, indicating that 
  they belong to a single ionized reflection component.  
The spectral shape, X--ray flux, and variability properties are very
similar in the \xmm\ and \suzaku\ observations, performed about 5
years apart. The overall X--ray spectrum and variability can be
described by a simple two--component model comprising a steep power
law continuum plus its ionised reflection off the inner accretion
disc. In this model, a rapidly rotating Kerr black hole
and a steep emissivity profile are required to describe the data.  The
simultaneous detection of broad relativistic Fe L and K lines in IRAS
13224-3809 follows that in another extreme NLS1 galaxy,
1H~0707--495. Although the data quality for
  \iras\ does not allow us to rule out competing models as in
  1H~0707--495, we show here that our reflection-based interpretation
  describes in a self--consistent manner the available data and points
  towards \iras\ being a very close relative of 1H~0707--495 in terms
  of both spectral and variability properties.
%The 
%particularly high Fe abundance and reflection strength in these two 
%objects is what make it possible to detect both relativistic lines so clearly. 
These results, together with those based on pure broad Fe K detections,  
are starting to unveil the processes taking place in the immediate vicinity of 
accreting radiatively efficient black holes.
\end{abstract}


\section{Introduction}

There is little doubt that narrow-line Seyfert 1 (NLS1) galaxies
provide an extreme view of the AGN phenomenon (e.g. Boller
\etal\ 1996, 2002; Brandt \etal\ 1997; Fabian \etal\ 2002, 2004; Gallo
2006; Gierlinski \etal\ 2008).  A strong soft excess below $\sim$2 keV
and significant X-ray variability are often seen, and with \xmm\ came
the discovery of high-energy spectral drop and curvature in some
objects (e.g. Boller \etal\ 2002, 2003).  The physical interpretation
is a topic of debate, and the suggestions are as extreme as the
objects themselves involving relativistically blurred reflection
(e.g. Fabian \etal\ 2002, 2004; Ponti et al. 2006) or ionised
absorption (Gierlinski \& Done 2004; Middleton \etal\ 2007), and
partial covering (Tanaka \etal\ 2004, 2005) that may be outflowing
(Gallo \etal\ 2004a).  A breakthrough was the discovery of broad iron
(Fe) L$\alpha$ emission and reverberation-like time lags in a long
observation of 1H~0707--495 (Fabian \etal\ 2009).  Confirmation of
this type of behaviour in other objects would bolster the importance
of NLS1s in understanding the innermost regions of AGN.

\iras\ ($z=0.0667$) along with 1H~0707--495 are perhaps the most
remarkable members of this class, displaying the above properties in
outstanding fashion.  In many ways \iras\ and 1H~0707--495 are very
similar objects, especially in the X--ray band.  They both show a high
energy Fe K drop around 7-8 keV implying either an outflowing partial
covering absorber or a large emission line (Boller et al. 2002;
Fabian et al. 2004; Boller et al. 2003). In both cases the material
producing the structure must be iron overabundant with respect to the
Sun. They also have large soft excesses (Ponti PhD thesis) with
sharp drops around 1 keV (Leighly 1999). Moreover they show a high
degree of variability indicating low black hole masses and high
accretion rates. %However,
%these objects are not unique and it has been suggested that  any NLS1 
%(perhaps any Seyfert 1) could at some times
%display such behaviour, that is, the underlying mechanism manifesting the
%extreme properties is the same in all NLS1 (Gallo 2006).

The motivation of this present work is to examine \iras\ in the light of
the discoveries from the new observations of 1H~0707--495 
(Fabian \etal\ 2009). To do so
we make use of a previously analysed \xmm\ observation (e.g. Boller
\etal\ 2003; Gallo \etal\ 2004b; Ponti et al. 2006) along with a new
\suzaku\ observation.  The purpose here is not to duplicate previous
efforts by fitting the spectra with partial-covering (or other
absorption models) and reflection models and comparing quality of
fits. Rather, our goal is to test the robustness of the reflection
model and show it to be a valid interpretation, similar to that used for
1H0707-495.

\section{Observation and data reduction}

\subsection{The \xmm\ observation}

\iras\ was observed with {\it XMM--Newton} on 2002 January 19 for
about 64~ks.  A detailed description of the observation and data
  analysis was presented in Boller et al. (2003) and Gallo et al. (2004),
  and we will briefly summarise details that are important to the
  analysis here.  Spectral files were created from the original ODF
using the SAS version 7.1.0. The EPIC pn and both MOS cameras were
operated in full--frame and large--window mode, respectively. The MOS1
and MOS2 spectra are affected by high background at energies greater
than 6 keV, so, we did not consider MOS data in the analysis, but used
these data only as a check of consistency. We analysed the RGS
  spectra, but the statistics hampers the detection of narrow
  absorption and/or emission features. Negligible pile up (checked
  with the {\sc epatplot} command) is present in the data and it does
  not affect the spectral results.  The total mean EPIC-pn count rate
  is 2.348 cts/s, and even the brightest flares are well
  below the maximum pile-up free count rate that can be achived in the
  full--frame EPIC pn science mode ($\sim$8 cts/s; Ehle et
  al. 2003). Consistent spectral shapes are obtained whether
  considering single events only or single and double events.  Thus,
  for this study, we selected the latter.  The source plus background
photons were extracted from a circular region with a radius of 32
arcsec. The background spectrum was extracted from source--free
regions on the same chip as the source. With the SAS commands
{\footnotesize ARFGEN} and {\footnotesize RMFGEN} ancillary and
response files were created.  After filtering periods of high
background (the biggest flare occurred approximately 20~ks after the
beginning of the observation and lasted for a few ks, see
Fig. \ref{fig:lcs}) we obtain net exposures of about 50~ks.  The total
observed flux is 5$\times$10$^{-13}$ ergs cm$^{-2}$ s$^{-1}$ and
2.3$\times$10$^{-12}$ ergs cm$^{-2}$ s$^{-1}$ in the 2-10 keV and
0.5-10 keV band, respectively. The count rate for the pn detector
  is 2.348 and 0.022 cts/s for the source and background. In our
  analysis the pn data has been considered in the 0.3-10 keV band.

\begin{figure} 
\rotatebox{270}
{\scalebox{0.34}{\includegraphics{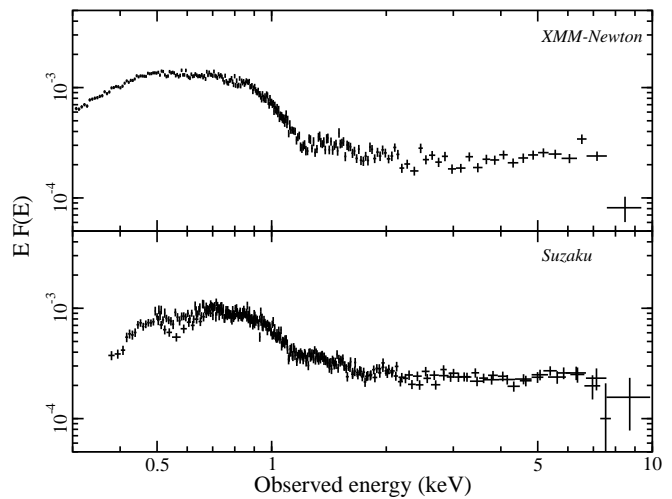}}}
\caption{Broadband spectra of \iras\ observed with \xmm\ pn (top panel) and
\suzaku\ XIS (lower panel).  The instrumental response is accounted for and
the axes are identical in the two panels to ease direct comparison.
The data are binned for appearance.
}
\label{fig:efespec}
\end{figure}
{The Optical Monitor (OM) data have been analysed starting from
  the Pipeline Products (PPS). Seven and 15 exposures were taken
  through the UVW2 filter and the UV Grism, respectively. We only
  present here the analysis of the images observed with the UVW2
  filter because of the low signal to noise of the grating
  observations. The OM UVW2 flux has been corrected for Milky Way
  (Galactic) reddening. The UV data have been converted into {\sc
    xspec} spectra files using {\sc flx2xps} utility, part of the {\sc
    ftools} package.  }

\subsection{The \suzaku\ observation}

\begin{figure*}
\includegraphics[width=0.46\textwidth,height=0.34\textwidth,angle=-0]{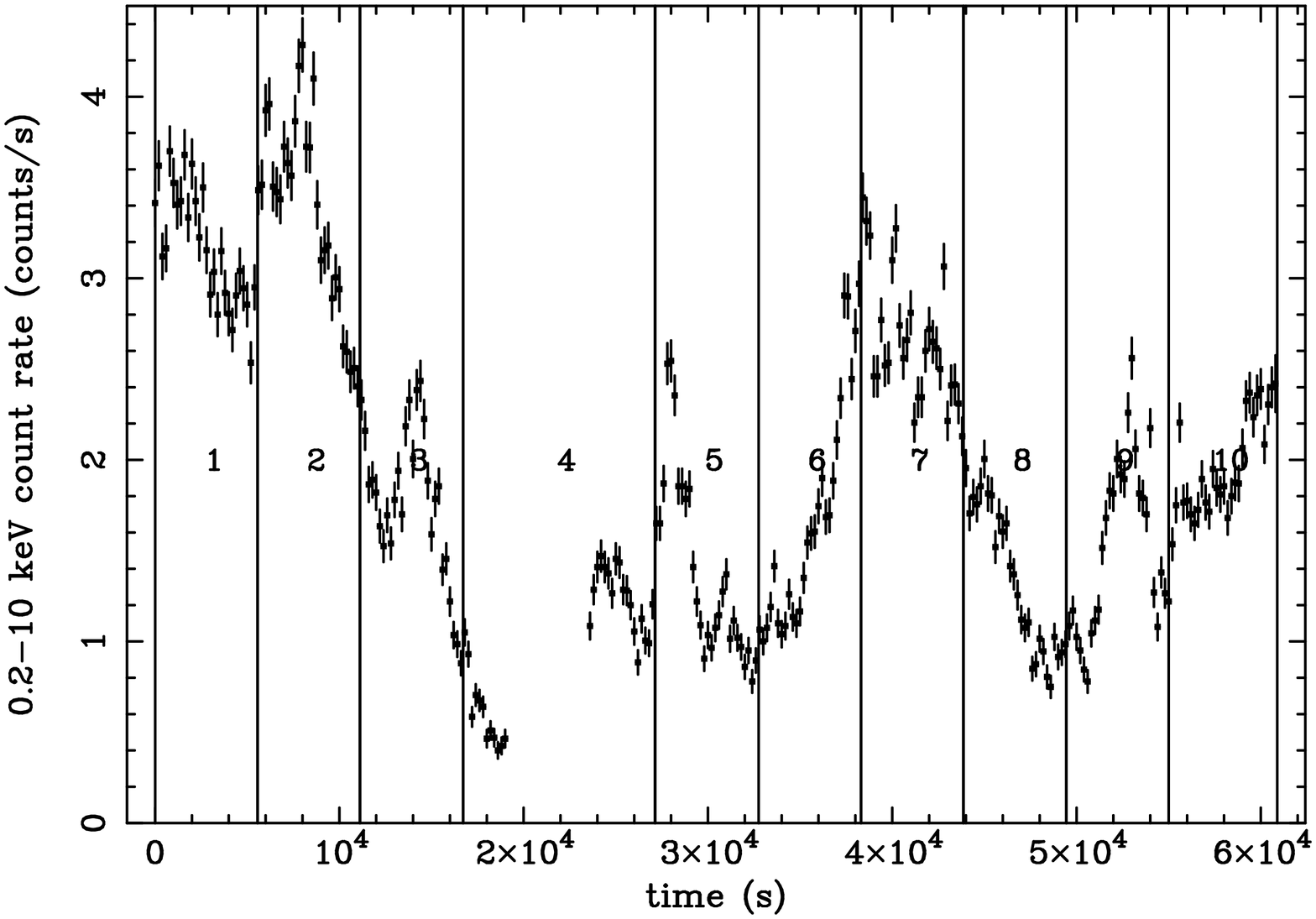}
\includegraphics[width=0.46\textwidth,height=0.34\textwidth,angle=-0]{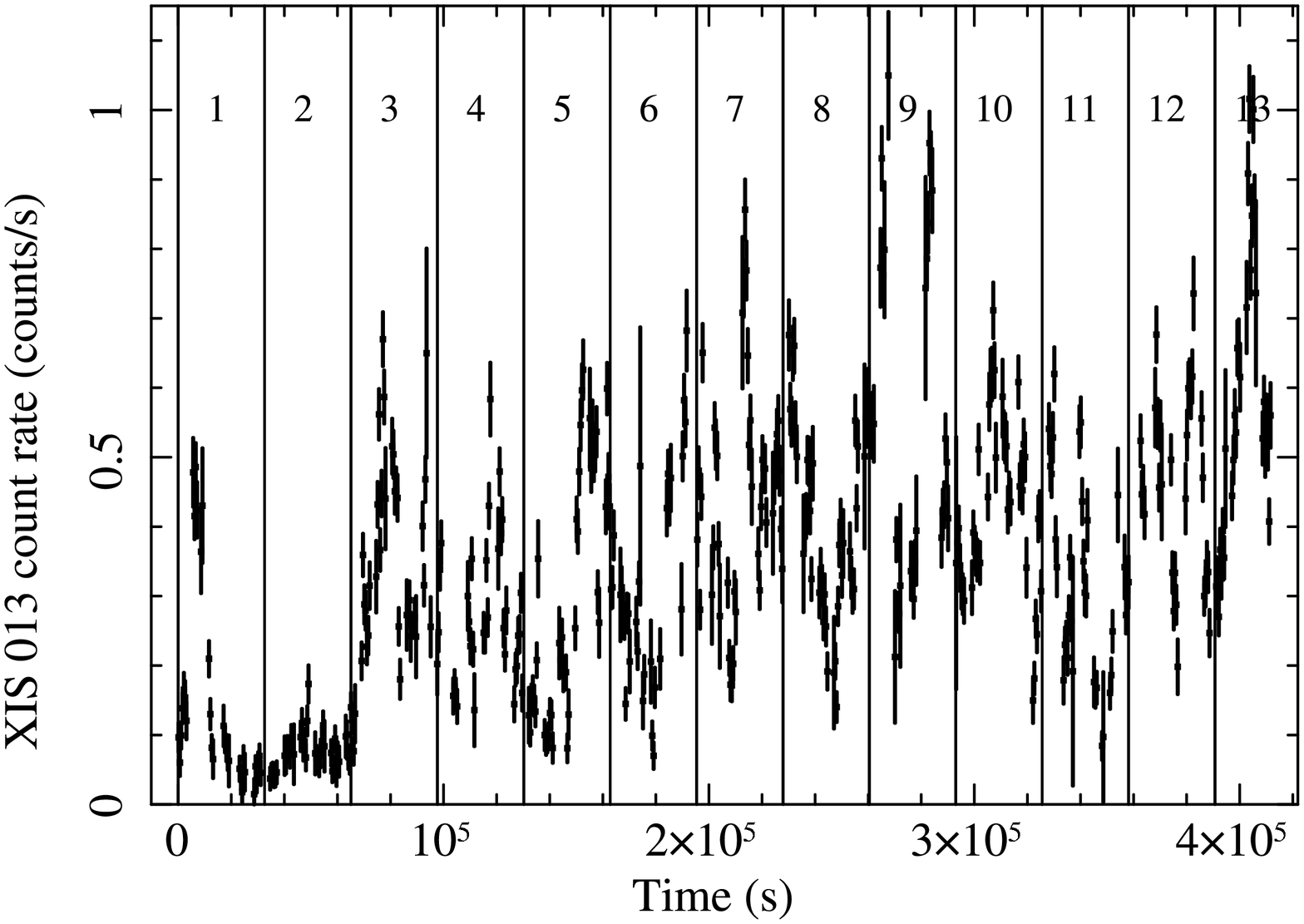}
\caption{Broadband light curves of \iras\ observed with \xmm\ pn in the 
$0.2-10$ keV band (left panel) and with the \suzaku\ FI-XIS in the 
$0.3-10$ keV band (right panel). The time intervals used for the time resolved 
spectral analysis are also shown.
}
\label{fig:lcs}
\end{figure*}
We observed \iras\ with the \suzaku\ X-ray Telescope. The 200 ks 
run started on 2007 January 26 at 05:48:34 and ended on 2007 
January 26 at 02:02:19 (UT).
During the time the two front-illuminated (FI) XIS detectors (XIS0 and XIS3)
and the single back-illuminated (BI) CCD (XIS1) operated in standard
pointing mode with the target centred on the detectors (XIS nominal pointing).
Event files were extracted from version 2.1.6.15 pipeline processing, using 
a large region of 260 arcsec, as suggested by the ABC Guide\footnote{http://heasarc.gsfc.nasa.gov/docs/suzaku/analysis/abc/}, 
in order to prevent
calibration problems. Background counts were extracted from annuli between 345
and 470 arcsec.
Spectra and light curves were generated using XSELECT.  Response matrices and
auxiliary response files were created using XISRMFGEN and XISSIMARFGEN 
(version 2007-05-14), respectively.  The spectra from the two FI CCDs
were examined separately and found to be consistent with each other within
uncertainties.  Consequently, the XIS0 and XIS3 data were merged to create a
single FI spectrum.  The HXD-PIN data were processed in the standard way,
but the source was not detected by the instrument.
The total observed flux is 7$\times$10$^{-13}$ ergs cm$^{-2}$ s$^{-1}$ and
2.4$\times$10$^{-12}$ ergs cm$^{-2}$ s$^{-1}$ in the 2-10 keV and 0.5-10 keV
band, respectively. The count rate for the XIS0, XIS1 and XIS3 detectors are 
0.156 cts/s (0.037), 0.299 cts/s (0.089) and 0.144 cts/s (0.031) for the source 
(background), respectively. In our analysis we considered the energy band 
0.4-10 keV for the XIS1 and 0.7-10 keV for the summed XIS0 plus XIS3. 
Above and below this range, deviations from reasonable models are found, 
indicative of calibration uncertainties. 

\subsection{Joint analysis}

In the subsequent spectral analysis, source spectra were grouped such that
each bin contained at least 20 counts and errors are quoted at the
90 per cent confidence level (\dchisq=2.7 for one interesting
parameter). Spectral fitting was performed using XSPEC 11.3.0.
All parameters are reported in the rest frame of the source unless specified
otherwise and all fits include Galactic 
absorption with the column density fixed at its nominal value
(5.3$\times$10$^{20}$ cm$^{-2}$; Kalberla et al. 2005).  

\section{A comparison of the \xmm\ and \suzaku\ data }

The \xmm\ and \suzaku\ observations are separated by nearly five
years, but nevertheless appear very similar both in terms of
brightness and shape (see Fig.~\ref{fig:efespec}).  This is
interesting considering the extreme spectral variability this object
experiences on short (hourly) time scales (e.g. Fig.~\ref{fig:lcs}).
The \xmm\ data show a clear feature in the Fe K band that can be
  reproduced by an absorption edge with energy E$=8.19^{+0.15}_{-0.20}$
  keV and $\tau=1.8^{+1.4}_{-0.8}$ (Boller et al. 2003).  The sharp
drop above 8 keV is difficult to detect in the \suzaku\ spectra as the
source starts to become background dominated at those energies,
however there are indications of it ($\tau=1.0^{+2.7}_{-0.8}$,
  $\Delta\chi^2 = 12.5$ for the addition of 2 new
  parameters\footnote{We observe that the drop is better detected
    ($\tau=0.7^{+0.4}_{-0.3}$, $\Delta\chi^2 = 16.9$ for the addition
    of 2 new parameters) when the spectra are extracted from a
    circular region with 2 arcmin radius (smaller than the standard
    4.3 arcmin radius), in this way reducing the impact of the
    background. Nevertheless, in order to avoid calibration problems
    of the instrument effective area, in the following analysis we
    consider the results obtained with the standard 260 arcsec
    extraction radius. }; see also lower panel of
Figure~\ref{fig:efespec}). Of certainty is that the feature has not
shifted strongly downward in energy.  The marginally detected $6.8$
keV emission line seen in the \xmm\ observation (Boller \etal\ 2003)
is not detected in the \suzaku\ spectrum with an upper limit on its
intensity of $6.7\times10^{-7}$ ph cm$^{-2}$ s$^{-1}$ and on the
equivalent width of 110 eV.

\begin{figure}
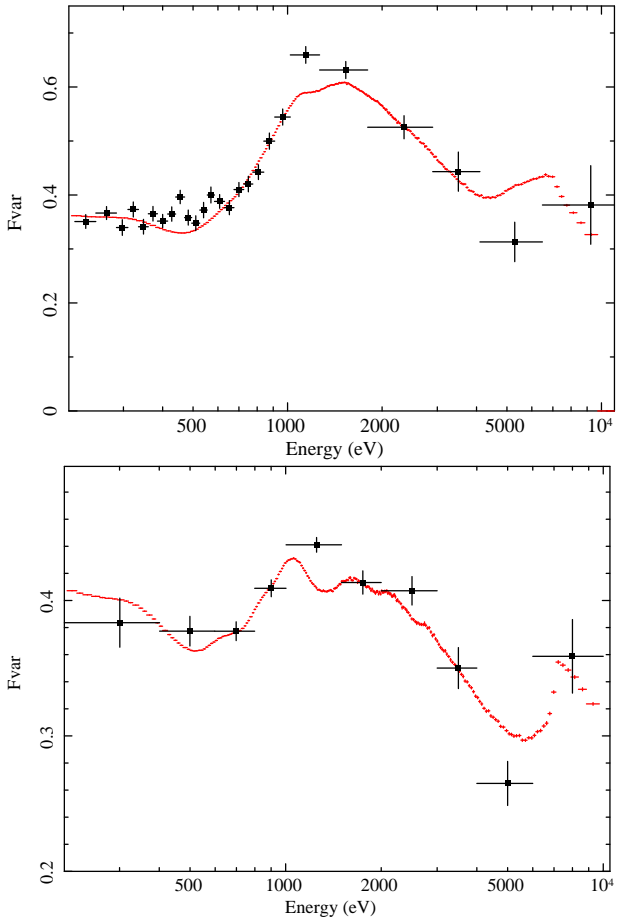
 
\rotatebox{270}
{\scalebox{0.34}{\includegraphics{RMS_5557_3.ps}}
\scalebox{0.34}{\includegraphics{RMS_Suzaku_3.ps}}}
\caption{RMS spectrum (see Ponti et al. 2004; Vaughan et al. 2003)
  from the \xmm\ (top panel) and \suzaku\ (lower panel) observations.
  Both are calculated using $\sim5$~ks bins, but the duration of the
  \suzaku\ observation is over five times longer. The red lines show
  the synthetic rms obtained from the best-fitting time resolved
  spectral variability study.  The simulated rms have been obtained
  starting from the best-fitting parameters with the power law plus
  ionised disc reflection model.}
\label{fig:rms}
\end{figure}

While the effective exposure was only three times longer during the
\suzaku\ observation, the duration of the observations was
substantially longer (over four days compared to three-quarters of a
day for the \xmm\ observation).  Consequently, there was a possibility
to examine variability on different time scales.  Despite this, the
\xmm\ and \suzaku\ light curves look similar (Figure~\ref{fig:lcs}),
displaying persistent variability with the largest outbursts being by
a factor of $\sim10$.  In addition, the shapes of the rms spectra (see
Figure~\ref{fig:rms}) are nearly identical with about a constant
amount of variability in the 0.2-0.7 keV band, a peak between 1 and 2
keV and lower variability at higher energies, with a drop around 5-6
keV. This does suggest that the processes dominating the variability
on short (hourly) time scales are likely the same as those which
dominate on daily time scales. The main difference between the two rms
spectra is tied to the amplitude of the drops. During the
\xmm\ observation the variability at low energy is about half that at
the peak, while during the \suzaku\ pointing the drop is only of the
order of 20 per cent.  This difference can be explained by the presence of a
stronger constant component during the \xmm\ observation (see \S 5.2
and Fig. 11).

Noteworthy, is that the huge amplitude (factor of $20-60$) 
outbursts recorded by $ROSAT$ (Boller \etal\ 1996) have never been seen since 
(Dewangan \etal\ 2002; Gallo \etal\ 2004b; this paper).

\section{UV-soft X-ray emission: a standard disc black body component?} 

\begin{figure} 
\includegraphics[width=0.2\textwidth,height=0.46\textwidth,angle=-90]{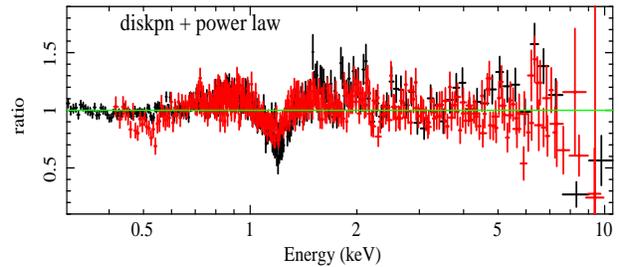}
\caption{Residuals from fitting a simple power law plus disc black
  body ({\sc diskpn}, Makishima et al. 1986; Gierlinski et al. 1999)
  model to the XMM-Newton (black) and Suzaku (both FI and BI in red).
  The disc black body emission broadly reproduce the soft excess
  emission, but leaves strong residuals at $\sim$1.3 keV.}
\label{ratdiskpn}
\end{figure}

We first note that a simple power law model fails to reproduce
  the X-ray source emission because of the large soft excess. We
  thus add the contribution from a disc black body emission component
  ({\sc diskpn}, Makishima et al. 1986; Gierlinski et al. 1999). Large
  residuals are still present at $\sim$1.3 keV and between 7 and 8 keV
  (see Fig. \ref{ratdiskpn}).  The black body component is, however,
  able to reproduce the soft excess emission.  The best fit inner disc
  temperatures (assuming an inner disc radius of 6 r$_g$) are
  T$_{BB}=0.148\pm0.004$ keV and T$_{BB}=0.152\pm0.007$ keV for the
  \xmm\ and \suzaku\ data, respectively. The best fit luminosity of
  the disc black body component is L$_{0.001-10 {\rm keV}}=9.7$ and
  $6.9\times10^{43}$ erg s$^{-1}$.

\begin{figure} 
\includegraphics[width=0.38\textwidth,height=0.48\textwidth,angle=-90]{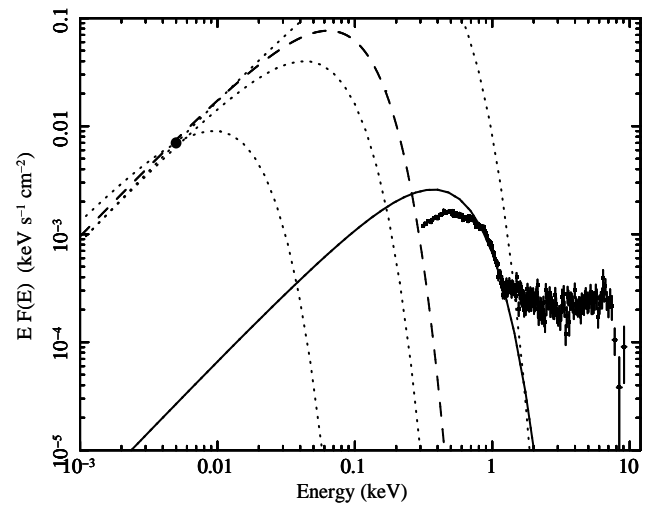}
\caption{\xmm\ EPIC pn and OM UVW2 data. The solid line shows the
  de-absorbed disc black body component that fits the soft X-ray
  data. The dotted lines show the ones fitting the UV data, assuming a
  black hole mass of 10$^8$, 10$^7$ and 10$^6$ M$_{\odot}$,
  respectively from bottom to top. The dashed line assumes a BH mass of 
  5.6$\times$10$^6$ M$_{\odot}$ (Wang et al. 2004).
  A unique disc black body component
  fails in reproducing both the UV and soft X-ray source emission. The
  most probable possibility is a BH mass slightly lower than 10$^7$
  M$_{\odot}$ producing the UV flux and a further emission component
  for the soft excess.}
\label{UV}
\end{figure}

Figure \ref{UV} shows the \xmm\ (EPIC-pn) and the simultaneous OM
UVW2\footnote{{ We also compared the OM data with the HST STIS,
  taken from Leighly \& Moore (2004). The two UV data-sets agree
  well with each other, despite being acquired several years apart. We
  also note that the STIS spectrum shows a flattening at lower
  energies, where the strong starburst component becomes important
  (Leighly 2004).}} spectra.  The solid line shows the un-absorbed disc
black body component ({\sc diskpn}) best fitting the soft X--ray
excess emission. Once that this component is extrapolated in the UV
band, it fails in reproducing the strong source UV emission.  In other
words, if the disc black body is responsible for the soft X--ray
excess, the UV flux is severely underestimated and cannot be accounted
for by the model.
%The soft
%X-ray emission can not be simply the high energy extrapolation of a
%hot disk black body emission.  }
%In order to get insight into the origin of the soft excess emission, we analysed 
%the OM data (UVW2 filter). }
%Initially the spectra from each epoch were fitted separately with a suite of ``physical''
%continuum models. a hot plasma Comptonizing soft photons ({\sc comptt}, 
%Titarchuk 1994), or In both cases similar 

We then fitted the UV and 2--10 keV data as described in
  Vasudevan \& Fabian (2007; 2009)\footnote{The authors use a 
  {\sc diskpn} model assuming an inner radius of 6 r$_g$.}. 
  The black hole mass of
  IRAS13224-3809 is not well known.  The fast and high degree of
  variability suggests a small BH mass (M$_{BH}<$0.7-1$\times10^7$
  M$_{\odot}$; Ponti et al. in prep.), while considerations based on
  the high source bolometric luminosity and the ultraviolet
  emission/absorption components suggest a M$_{BH}$ from about
  $6\times10^6$ to $10^8$ M$_{\odot}$ (Leighly \& Moore 2004; Leighly
  2004; Wang et al. 2004). To study the impact of the uncertainty
  associated with the poor knowledge of the BH mass, we consider three cases 
  with BH masses of $10^6$, $10^7$ and $10^8$ M$_{\odot}$.  The
  dotted lines in Fig. \ref{UV} show the best fit model reproducing
  the UV and 2-10 keV spectra\footnote{Similar results are obtained
    considering the power law emission in the 2-3 and 8-10 keV bands
    only.} for the different BH masses. If a BH mass of 10$^7$ or
  10$^8$ M$_{\odot}$ is assumed, it is not possible to reproduce the
  soft excess with a standard disc black body component because the
  disc temperature is too low.  On the other hand a lower BH mass, in
  order to fit the UV data, would imply an un-physically high
  Eddignton accretion rate (L/L$_{Edd}>100$ for a 10$^6$ M$_{\odot}$,
  L/L$_{Edd}\sim1$ for a 10$^7$ M$_{\odot}$).
  
  The dashed line in Fig. \ref{UV} shows the best fit model assuming 
  a BH mass of 5.6$\times$10$^6$ M$_{\odot}$ (Wang et al. 2004).
  In this case a super-Eddington accretion is required ($\eta\sim4$), 
  however a still hotter black body temperature is required to fit the 
  soft excess.

Modification to the disc black body emission due to 
atomic processes (in particular hydrogen and helium) are an 
interesting possibility. In fact in AGN, due to the low free-free and 
bound-free opacities in the soft X-ray regime (that allow the diffusion 
of photons from large Thomson depths below the disc surface), a high 
energy tail is generated in addition to the standard UV disk black body 
shape. However, this component, having a too low intensity and 
"temperature" (the expected maximum temperature being 50-60 eV), 
can not explain the soft excess emission of IRAS 13224-3809 
(Ross et al. 1992).

%Nevertheless, this process with the contribution 
%from hard X-ray irradiation (reflection) may explain the soft excess 
%emission (the luminosity of the disk black body emission extrapolated 
%from the soft X-ray is $\sim10^{44}$ erg s$^{-1}$, 3-5 \% of the UV 
%disc black body emission). 

Alternatively, as already explored by Gierlinski \& Done (2004), a
second lower temperature Comptonization medium (with lower energy
electrons compared to the ones producing the power law emission above
2 keV) may fit the soft excess spectral shape. IRAS13224-3809 makes no
exception.  Nevertheless, we obtain (fitting the soft excess with a
Comptonization component {\sc compST} in Xspec) an electron
temperature T$_{soft}\sim0.13$ keV and an optical depth $\tau\sim90$.
These values fall within the narrow range of values observed by
Gierlinski \& Done (2004) analysing the soft excess of a large sample
of AGN.  They observed clustering of all the measured temperatures and
optical depths to a narrow range of values that poses serious doubts
on the second Comptonisation hypothesis for the soft excess
(Gierlinski \& Done 2004) and brought the authors to exclude this
hypothesis. Thus, although the model can explain the soft excess in
any single source, it seems to fail when the big picture is
considered.  Moreover, we stress that although soft state spectra of
GBH can be modelled by two media of Comptonised electrons, the
observed optical depth, in that case, is definitely smaller than
$\tau\sim90$ (as required here) resulting in a far smoother
shape of the soft excess, than the one observed here. 

The main conclusion drawn from this exercise is that a unique disc
black body component can not reproduce both the UV and soft X-ray
data. The most likely situation is then that a thermal accretion disc
component is responsible for most (if not all) of the UV emission,
while the soft X-ray emission has a different physical origin.

 This is in agreement with the observation that in AGN, in
  general, when the soft excess is fitted with a disc black body
  component, the resulting temperature is observed to be too high and
  with a too small scatter, compared with expectations (only in the
  smaller mass and higher accretion rate AGN the Wien tail of the disc
  emission is expected to be observed in the soft X-ray band). For
  this reason many works suggest that the real origin is tied to
  ionised absorbing/reflecting material (Gierlinski \& Done 2004;
  Crummy et al. 2006). Furthermore, in observations allowing us to
  study the X--ray variability of this putative thermal component, the
  expected $L\propto T^4$ blackbody relationship is not recovered,
  casting further doubts on this interpretation (e.g. Ponti et
  al. 2006).

%Modification to the disc black body emission due to 
%atomic processes are an interesting possibility. In fact in AGN, 
%due to the low free-free and bound-free opacities in the soft X-ray regime 
%(that allow the diffusion of photons from large Thomson depths below the 
%disc surface), a high energy tail is generated in addition to the standard 
%UV disk black body shape. This component alone can not explain the 
%soft excess emission (Ross et al. 1992), the expected maximum temperature 
%being 50-60 eV. Nevertheless, this process with the contribution 
%from hard X-ray irradiation (reflection) may explain the soft excess 
%emission (the luminosity of the disk black body emission extrapolated 
%from the soft X-ray is $\sim10^{44}$ erg s$^{-1}$, 3-5 \% of the UV 
%disc black body emission). 

\section{Ionised absorption}

Alternatively to the disc black body emission (or cold Comptonization)
  thermal models, the entire soft excess
may be attributed to the effect of partially ionized gas in the line
of sight, absorbing the X--ray continuum which would be then
intrinsically more intense and significantly steeper than observed in
the 2--10 keV range.  As shown by Boller et al. (2003), the high energy
drop at about 8 keV can indeed be produced by a neutral outflowing
absorber partially covering the source.  Iron overabundance is
required to fit the feature. Alternatively, a partial covering ionised
absorber, similar to the one suggested for 1H0707-495 (Done et
al. 2007), may explain the high energy feature of IRAS13224-3809,
although the edge seems too sharp to be produced by partially ionized
gas.

\begin{figure} 
\includegraphics[width=0.25\textwidth,height=0.47\textwidth,angle=-90]{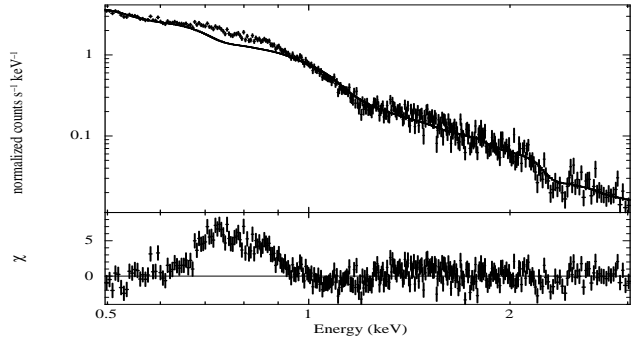}
\caption{The source spectrum fitted with a black body plus power 
law model absorbed by two partial covering absorber components. 
To attribute the $\sim$1.3 keV feature to Fe L absorption requires strong
Fe M absorption at $\sim0.8$ keV that is not seen in the spectra. 
}
\label{Abdu}
\end{figure}

Nevertheless, we note that to fit the source spectrum below 2 keV two
additional components are needed, an emission one (either a disc black
body or a second power law) for the soft excess and another different
absorption component to fit the $\sim$1.3 keV structure.  We thus fit
the spectrum with a power law plus a disc black body emission absorbed
by two ionised partial covering components\footnote{We used the
  same grid of models as Zoghbi et al. (2010), generated through the
  XSTAR photoionisation code (Kallman et al. 1996). The models have
  abundances fixed to Solar and the illuminating flux has a steep
  power law index of 3. The fitting parameters are the absorber column
  density, ionisation parameter, covering factor and Fe abundance.}.
With one absorber roughly reproducing the high--energy drop as in
Boller et al. (2003), while the second absorber can reproduce 
the low energy structure only if significant blue-shifting 
($\sigma\sim$0.15-0.3 c) is allowed (fitting the structure with either 
O {\sc VIII} or Fe M UTA).
No good fit was found with lower outflow velocities.
The main reason is
that the model tries to fit the apparent drop below 1 keV with an
L-shell iron edge, requiring high iron abundances. However, as
Fig. \ref{Abdu} shows, this predicts a drop between 0.7 and 0.9 keV
due to the M-shell unresolved transition array (UTA) (Fabian et
al. 2009; Zoghbi et al. 2010). This is a blend of numerous absorption
lines arising from the photo-excitation of ions FeI -FeXVI mainly
produced by 2p-3d transitions (Behar et al. 2001).
Moreover the absorption-dominated model still requires a blackbody
to model the soft excess.

As mentioned in $\S$1, the purpose of this paper is not to
  fit the complex X--ray spectrum of \iras\ with all possible models
  and compare the best--fitting $\chi^2$. Our goal is to seek the
  simplest possible solution that can explain both the spectral and
  variability properties with a minimal set of assumptions. We suspect
  that combining the soft blackbody and the power law with further
  layers of partial covering absorbers with different column
  densities, ionization stages, covering factors, and with unrelated
  outflowing velocities could easily provide a fair description of the
  X--ray spectrum. However, the model would be poorly constrained
  because of its complexity and would not improve our understanding of
  the source. We thus consider here a different interpretation of the
  X--ray properties of \iras\ with the goal of describing the data
  with a minimal set of physically connected model components.

\section{Signatures of disc reflection?}

The X-ray spectrum of 1H0707-495 shows similar features at $\sim$1.3
keV and in the Fe K band. Fabian et al. (2009), Zoghbi et al. (2010)
and the results of the detailed analysis of the RGS spectra confirm
that the origin of the 1 keV feature is not due to absorption (Blustin
\& Fabian 2009). On the other hand these structures and the timing
properties of the source, are consistent with being produced by
relativistic Fe L and K lines. Thus, we investigate here the
consistency of the source emission with the reflection interpretation
and its similarity with 1H0707-495. 

We thus reproduce the phenomenology of the continuum emission as in
Fabian et al. (2009) with a power law plus disc black body emission.
This is a crude approximation of the continuum emission. In
particular, as shown in $\S$4, the black body component is not
associated with the disc emission, being in this interpretation
associated with the reflection continuum.  This study is important to
check the similarity with 1H0707-495. We observe that, as in the case
of 1H0707-495, a significant improvement is achieved for all fits when
two broad disc lines with energies of $\sim1$ keV and $E>6$ keV
(depending on the continuum model) are added to the
continuum\footnote{A $\Delta\chi^2$ of 385 and 79 is observed when a
  broad Fe L and Fe K line is respectively added to the source
  spectrum. The strong statistical variation indicate the importance
  of these features in the spectrum.}.  The lines are broadened by the
motion of the emitting material on the surface of an accretion disc
and by the relativistic effects present in the vicinity of a black
hole (in Xspec {\sc Laor} profile, Laor 1991). The model parameters
are the emitting energy (E), intensity (norm), disc inner and outer
radii (r$_{in}$, r$_{out}$), disc inclination (i) and power law disc
emissivity index (q).  The energies are broadly consistent with those
expected from \fel\ and \fek\ for a common ionisation parameter
($log(\xi)\sim2-3.5$; Kallman 1995; Kallman et al. 1996; 2004).
It is also worth noting that the high- and low-energy Laor profiles are 
consistent with identical blurring parameters (see Tab. \ref{Laors}).  

\begin{figure} 
%\rotatebox{270}
%{\scalebox{0.34}{\includegraphics{iron.ps}}}
%\hspace{-0.6cm}
 \includegraphics[width=0.495\textwidth,height=0.39\textwidth]{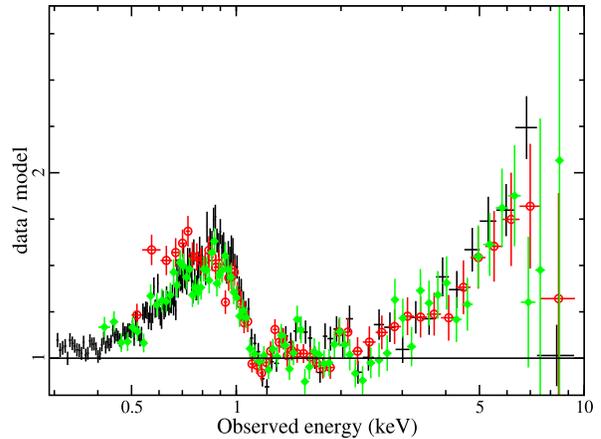}
%\vspace{-7.05cm}
%\hspace{3cm}
%\includegraphics[width=0.46\textwidth,height=0.3\textwidth]{ratio_FeL_K.ps}
%\vspace{4cm}
\caption{%{\it Upper panel} 
  Residuals from fitting a black body plus power law continuum 
  with two Laor profiles then setting the
  normalisation of the Laor profiles to zero. Black (pn), red (FI) and
  green (BI) models are shown. This is clearly a model--dependent way
  of showing the residulas (see Fig.~\ref{ratdiskpn} for comparison)
  but we believe it is useful to show it here for comparison with a
  similar figure presented in Fabian et al. (2009) for the much better
  statistics case of 1H~0707--495.)
%{\it Lower panel} Superposition of the profiles of the Fe L 
%(black filled squares) and Fe K (red open circles). Only the pn data are 
%shown for clarity reasons.
}
\label{fig:iron}
\end{figure}

Fig.~\ref{fig:iron} shows the residuals to black body plus power law
continuum fits after setting the normalisation of the Laor profiles to
zero, clearly showing the large soft and hard residuals, both
reminiscent of broad and skewed emission line profiles, as expected in
the case of relativistic disc lines.  The resemblance with 1H0707-495
is impressive (Fabian et al. 2009).  We observe that the relativistic
parameters associated with the low and high energy lines are
consistent with each other (see Tab. 1).  
This suggests that the
  broadening of the low--energy feature due to the different lines of
  the Fe L complex and to other soft emission lines is likely
  negligible (although it is likely that these effects are one of the
  the causes of the inferred steep disc emissivity profile).

Given the consistency of the relativistic parameters for the two
features, to better constrain the two
broad line hypothesis, we impose the same disc inclination, emissivity
index and inner disc radius. The line energies are still consistent
with Fe L and K emission (see lower panel of Tab. 1). Moreover, the
normalization of the Fe L and K are in the ratio $30\pm11$, consistent
with expectations from an ionised disc reflection (Ross \& Fabian
2005; see also \S 5.2.1).
The equivalent width of the Fe L (Fe K) line is 0.2 keV (2.8 keV) and
0.07 keV (0.5 keV) during the \xmm\ and \suzaku\ observations,
respectively.  We point out that the inferred value of the Fe K line
EW is extreme.  Assuming Fe solar abundance, the Fe K line EW with
respect to its own reflection continuum is not expected to be much larger
than about 1 keV for spectra with $\Gamma \geq 2$. A 5-10$\times$Solar
Fe abundance produces a factor ~2-3 increase in the line EW (George \&
Fabian 1991) which is in principle able to account for the observed EW
if the spectrum is reflection-dominated. However, it should be pointed
out that the reflection continuum in the Fe K band is not a simple
power law and has a skewed shape dropping sharply at the Fe edge.
Hence a simple broad line plus power law fit is very likely to
overestimate the line EW because a large part of the reflection
continuum is erroneously accounted for by the broad line. We conclude
by pointing out that the numerical results and parameters obtained
with the phenomenological model presented above must be taken with
caution and the model is presented here only as a guide for the more
sophisticated and realistic reflection models that will be discussed
in the following sections. We stress that, as in 1H0707-495, the two
main features of the X-ray spectrum, the sharp drops at $\sim$1.1-1.2
keV and $\sim$8.2 keV (observed frame), can be reproduced by two
emission lines with similar ratios in the two objects.
%A unique reflection component
%could reproduce the main features of the spectrum while at least two
%different absorption components are required to fit the two features
%and, moreover, an absorption-dominated interpretation is not
%consistent with the lack of Fe M absorption in the data.

We note that, even if still too high, the inclusion in the model
  of the Fe L emission significantly lowers the temperature of the too
  hot disc component required to fit the soft excess, going from
  T$_{BB}=0.148-0.152$ keV to T$_{BB}=0.118-0.122$ keV. Basically, the
  Fe L feature well reproduces the high energy part of the
  soft excess emission. The inclusion of all the lines present in a
  disc reflection spectrum may explain the whole soft X-ray emission
  without having to invoke a thermal disc component.
\begin{table}
\scriptsize
\begin{tabular}{c cc cc c}
\hline 
\hline 
               & \xmm\        &                                & \suzaku\ \\
\hline 
              & T$_{BB}$  & $\Gamma$  & T$_{BB}$  & $\Gamma$       & \\
               & $118\pm4$ & $2.7\pm0.1$ &$122\pm10$& $2.5\pm0.1$        \\ 
\hline                
\multicolumn{5}{l}{Independent disc parameters tied between Fe L and Fe K } \\ \\
\hline
	        & Fe L                            & Fe K                      & Fe L                            & Fe K                        \\
E            &$1.14^{+0.05}_{-0.09}$&$6.3\pm0.5$          &$1.07^{+0.10}_{-0.11}$&$6.7\pm0.5$            \\
q            & $7.1^{+1.7}_{-0.3}$     &$6.5\pm1.1$           & $7.3^{+1.5}_{-0.4}$     &$7.8\pm2.2$            \\
R$_{in}$& $1.58^{+0.3}_{-0.12}$ &$1.3^{+0.5}_{-0.1}$ & $1.6^{+0.4}_{-0.2}$    &$1.3^{+0.2}_{-0.1}$ \\
Inc         & $40^{+11}_{-18}$        &$67^{+10}_{-30}$    & $48^{+6}_{-20}$         &$60^{+10}_{-20}$    \\
norm     & $93\pm5$                    &$1.8^{+0.6}_{-0.3}$  & $37^{+12}_{-3}$        &$1.8^{+0.4}_{-0.6}$  \\
\hline
\multicolumn{5}{l}{Disc parameters tied between Fe L and Fe K } \\ \\
\hline
E           &$0.92\pm0.02$       &$6.68\pm0.3$        & $0.93\pm0.04$    & $6.5\pm$0.2 \\
q           &$8.3^{+1.8}_{-0.4}$&                              & 8.1$_{-1.5}^{+0.8}$  \\
R$_{in}$&$1.34\pm0.07   $   &                               & $1.38\pm0.05$ \\
Inc         &$66^{+18}_{-12}$  &                               & $66\pm2$ \\
norm     &$55^{+8}_{-7}$      &$1.8^{+0.5}_{-0.3}$& $35_{-8}^{+17} $ & $1.2_{-0.3}^{+0.4} $\\
\hline
\end{tabular}
\caption{Best fit results of the \xmm\ and \suzaku\ data with the parametric model composed 
of power law plus disc black body emission ({\sc diskpn} assuming an inner disc radius of 6 r$_g$) 
and two lines broadened by a relativistic profile (Laor et al. 1991). The two broad lines have 
energies consistent with Fe L and Fe K. The disc black body temperature and normalisation 
are in units of eV. The lines energies, inner radii, inclinations and normalisation in keV, 
gravitational radii, degrees and $10^{-5}$ ph cm$^{-2}$ s$^{-1}$ in the line.
Even imposing the same parameters for the relativistic profiles the best fit energies are still 
consistent with emission from Fe K and Fe L and the line ratio with the expected one 
(Ross \& Fabian 2005).}
\label{Laors}
\end{table} 

\subsection{Self-consistent ionised disc reflection}

\begin{table*}
\scriptsize
\begin{tabular}{cc cc cc cc cc}
\hline 
\multicolumn{9}{l}{ionised disc reflection + power law: ({\sc wabs*(kdblur*reflionz+powerlaw)})} \\ \\
\hline
               & $\Gamma$     & Abun$_{Fe}$ & $\xi$                         & q                                  & r$_{in}$                         & r$_{out}$ & inc.             & Flux$_{3-10}$ \\
  \xmm    & 2.45$\pm$0.01&  5$\pm$0.2   & 537$^{+16}_{-25}$  & 5.93$^{+0.14}_{-0.1}$ & 1.235$^{+0.14}_{-0}$     & 400          & 52$\pm$1  & 3$\times 10^{-12}$ \\
\suzaku  & 2.38$\pm$0.02 &  -                  & 349$\pm18$            & 4.85$^{+0.25}_{-0.2}$ & 1.235     & -               & -  & 3$\times 10^{-12}$ \\
\hline 
%\multicolumn{12}{l}{ionised disc reflection + power law + disk black body: ({\sc wabs*(kdblur*reflionz+powerlaw+diskpn)})} \\ \\
%\hline
%               & $\Gamma$ & Abun$_{Fe}$ & $\xi$ & q & r$_{in}$ & r$_{out}$ & inc. & T$_{BB}$ & A$_{BB}$ & Flux$_{3-10}$ & \\
%  \xmm    & \\
%\suzaku  & \\
\end{tabular}
\caption{Best fit results of the \xmm\ and \suzaku\ data with the parametric model 
composed of power law plus ionised disc relativistic reflection. We stress again 
that these errors take into account only statistical uncertainties.}
\label{reflion}
\end{table*}
An ionised reflection model (Ross \& Fabian 2005) modified by the
relativistic effects occurring on an accretion disc in the proximity
of the black hole (e.g. Fabian \et 2004; Crummy \et 2006; Ponti \et
2006; Petrucci et al. 2007; Larsson et al. 2008; Ponti et al. 2009)
could potentially reproduce the mean spectrum.  As shown in $\S$4, the
disc black body emission can explain the UV emission, but fails in
reproducing the soft excess. A prominent broad Fe L line, alone, can
fit the high energy tail of the soft excess, but still requires
  the presence of a blackbody disc component (although a slightly
  colder one). The inclusion of the
emission lines associated with the other elements may entirely explain
the soft excess emission. For this reason we fit the spectrum with a
simple power law model plus ionised reflection.

\begin{figure} 
\rotatebox{270}
%%%{\scalebox{0.34}{\includegraphics{refl_efe.ps}}
%%%\scalebox{0.34}{\includegraphics{refres.ps}}}
{\scalebox{0.385}{\includegraphics{refl_efe.ps}}
\scalebox{0.34}{\includegraphics{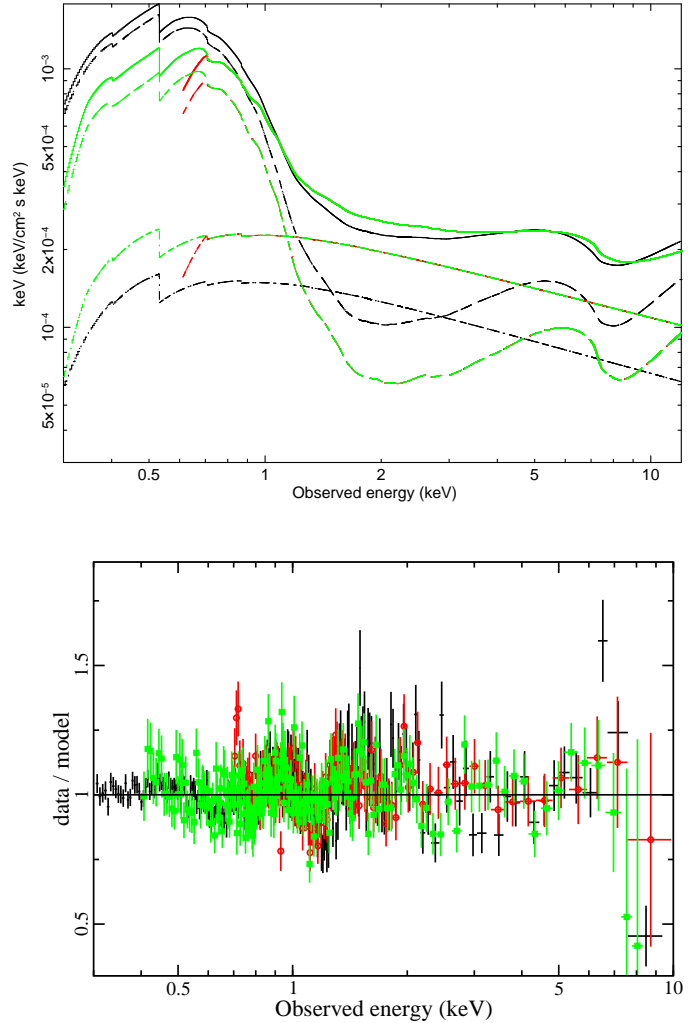}}}
\caption{Top panel: Blurred reflection plus power law model components.
Black (pn), red (FI) and green (BI) models are shown.  Lower panel:
The residuals from the above fit (colour code as before). 
}
\label{fig:refres}
\end{figure}

The \xmm\ and \suzaku\ spectra are fitted simultaneously and in a
self-consistent manner. Parameters that are not expected to change
over time scales of a few years, for example: neutral absorption (but
see Gallo \et 2007); disc inclination; and iron abundance, are fixed
between the data.  All other parameters are free to vary. The best fit
models and the residuals from such a fit are shown in Table
\ref{reflion} and in Figure~\ref{fig:refres}.
 
The model fits show a disc inclination of $\sim52$ degrees; an inner
and outer radius consistent with $1.235$ and $400$ $r_g$ ($1r_g =
GM/c^2$) respectively, and an iron abundance $\sim5$ times solar
(Morrison \& McCammon 1983).  The disc emissivity profile is
comparable at both epochs ($\sim5$) as is the power law continuum
photon index ($\Gamma\approx2.4$).  The ionisation parameter (defined
as: $\xi = \frac{4 \pi F}{n}$; Tarter, Tucker \& Salpeter 1969, where
F is the total illuminating flux and n is the hydrogen number density)
changes only slightly between the two epochs: $\sim500$ erg cm
s$^{-1}$ during the \xmm\ observation compared to $\sim340$ erg cm
s$^{-1}$ at the later epoch. 

Assuming that the reflecting disc is truncated at the innermost stable
circular orbit (ISCO) around the black hole, the inner disc radius
depends only on the black hole spin (see i.e. Reynolds \& Fabian
2008). The breadth of the lines in IRAS13224-3809 requires an inner
radius smaller than $\sim$2 gravitational radii (90 \% confidence
level), indicating a spin parameter higher than $a>0.94$.

The average $0.3-10$ keV flux at the two
epochs is also comparable ($\sim3\times10^{-12}$\ergcms).  The main
difference between the two observations is the ratio between the flux
of the direct and reflected component, with the \xmm\ spectrum being
almost entirely reflection dominated (this is in agreement with the
observed line equivalent width, see \S 4.1).  The $10-50$ keV band is
reflection dominated with a predicted flux of
$\sim5\times10^{-13}$\ergcms, and consistent with the lack of
detection in the HXD.  The fit is reasonably good ($\chi^{2}_{\nu}/dof
= 1.13/1536$) compared with other two-component continuum models,
although some residuals are still present (Figure~\ref{fig:refres}).

The advantage of the above interpretation is that the \fel\ and \fek\ 
are predicted features of the reflection model. Moreover the model reproduces 
the soft excess without the need to invoke a disc black body  
(negligible contribution is due to the extrapolation in the soft X-ray range 
of the disc black body emission fitting the UV data).
In this case the strong UV emission would be due to the disc emission, 
while the soft X-ray by ionised reflection. Even forcing an additional hot 
disc black body component to the best fit model, we obtain a hot disc 
temperature of about 0.12 keV. This seems too high even for being the 
hard tail of the disc black body emission due to the low free-free and 
bound-free opacities present in the soft X-ray domain (Ross et al. 1992). 

\section{Source spectral variability}

The \xmm\ and \suzaku\ broadband light curves of \iras\ (see Figure
\ref{fig:lcs}) exhibit large amplitude count rate variations, typical
of this source.  We explore here the time--resolved spectroscopy on
the shortest possible time scale (set by requiring good quality
time-resolved spectra).  The \xmm\ and \suzaku\ source light curves
were sliced into 10 and 13 time intervals, respectively (see Figure
\ref{fig:lcs}), and spectra were created for each time bin.

\subsection{Phenomenological model}

In order to gain insight on the source spectral variability and,
  in particular, to detail the variations of the Fe K and L lines, we
  fit the spectra with a phenomenological model composing of a power
  law plus a low energy disc black body continuum and two prominent
  broad lines representing the Fe K and Fe L emission. The disc black
  body component is required in order to approximately reproduce the
  effect of the ionised reflection continuum.
The disc black body temperature is fixed to the best fit values from
the mean spectra (T$_{BB}$=118 and 122 eV for \xmm\ and \suzaku,
respectively).  Similarly, the parameters of the relativistic line
profiles and their emission energies (see Tab. 1) were fixed. The
model reproduces well the spectra at all flux levels.

We observe that IRAS13224-3809 shows a correlation between the flux
and the power law spectral index, with the spectrum becoming steeper
with increasing flux.  The spectral index shows a large variation
during both observations, going from $\Gamma=1.7$ to about 3. This
behaviour has been observed in many other sources (see e.g. Fig. 10 of
Ponti et al. 2006) and it can be: i) either intrinsic due to
  changes in the parameters of the Comptonising phase, producing
  variations of the observed power law spectral index; ii) or spurious
  being due to the variations of the absorbing material that is
  dominating the spectral shape; iii) or due to a two component model
  comprising a steep power law varying in normalisation only and a
  harder and more constant component (McHardy, Papadakis \& Uttley
  1998; Shih, Iwasawa \& Fabian 2002).  The lack of a strong narrow
component in the Fe K line (the upper limit on a neutral line
  being EW $\sim30$ eV both for \xmm\ and \suzaku\ and on a ionised being
  EW=$120^{+140}_{-117}$ eV, E=$6.85^{+0.2}_{-0.1}$ keV for \xmm\ and
  $<110$ eV for \suzaku), excludes a strong contribution from a
constant reflection component from distant material. Thus, in this
latter case the constant reflection must come from the inner disc.

\begin{figure} 
 \includegraphics[width=0.46\textwidth,height=0.34\textwidth]{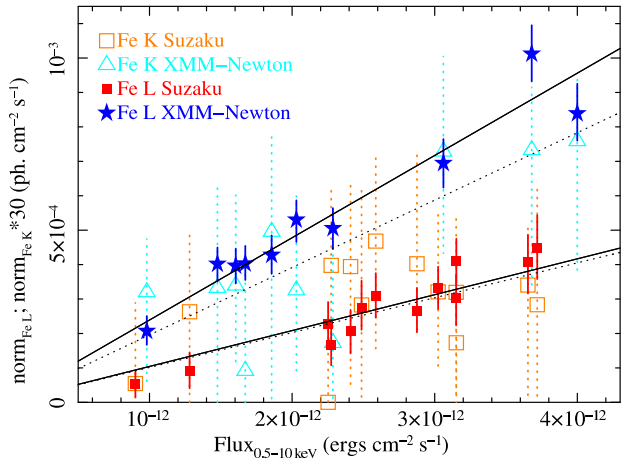}
\caption{
%{\it (Left panel)} Best fit power law spectral index ($\Gamma$) as a
%function of the 0.5-10 keV band source flux. A clear correlation is
%present, with the source spectrum becoming steeper with flux. The
%best fit values of the \xmm\ and Suzaku data are show with blue
%filled stars and red filled squares, respectively. {\it (Right
%panel)}
The filled blue and open light blue stars show the Fe L and Fe K
measurements from the \xmm\ observation, while the filled red and open
orange squares show the measurements from the \suzaku\ (the Fe K
normalisation has been multiplied for 30).  Filled symbols are for Fe
L and open ones for Fe K.  The solid and dotted lines show the best
fit linear relations norm$_{Fe L}$=a$\times$Flux$_{0.5-10 keV}$ in the
case of Fe L (a=10.3$^{+1.1}_{-1.2}\times$10$^7$;
10$^{+5}_{-4}\times$10$^7$ ph. ergs$^{-1}$) and Fe K
(a=23.9$\pm1.5\times$10$^7$; 20$\pm5\times$10$^7$ ph.  ergs$^{-1}$),
respectively.  The intensities of both broad lines are correlated with
the continuum at each flux level.  At the same flux, the Fe lines are
significantly more intense during the \xmm\ observation.
Nevertheless, the ratio between the Fe L and Fe K is consistent with
being constant not only during each single observation, but also
between the two observations taken 5 years apart. This suggests a link
between the two spectral features. The Fe L/Fe K ratio is consistent
with reflection from an ionised disc (Ross \& Fabian 2005).  }
\label{sv2laors}
\end{figure}

Figure \ref{sv2laors} shows the variations in the normalisation of Fe
L and Fe K as a function of flux (for display purposes the intensity
of Fe K has been multiplied by a factor of 30). The best fit values
for the \xmm\ observation are shown with filled blue and open light
blue stars for Fe L and Fe K, respectively, while the \suzaku\ ones
are shown with the filled red and open orange squares.  The
intensities of the two lines are well fitted by linear relations
during both observations\footnote{Only for the Fe K line during
  the \suzaku\ observation a fit with a constant gives comparable
  results to the fit with a linear relation ($\chi^2=3.8$ and 3.3,
  respectively), while during the \xmm\ pointing a fit with a linear
  relation provides a better fit $\Delta\chi^2=6.9$ (for the same dof,
  $\chi^2=9.6$ and 2.7, respectively). The Fe L is always better
  fitted with a linear relation $\Delta\chi^2=98.2$ and 105 (for the
  same dof, $\chi^2=57.3$ and 7.2 during the \suzaku\ and
  $\chi^2=113.0$ and 7.2, during the \xmm\ observation).}.  In fact,
once the Fe K intensity is multiplied for a factor of 30 (see Tab. 1),
we obtain the best fit linear relation for the Fe L and Fe K
  lines are: norm$_{Fe L}$=a$\times$Flux$_{0.5-10 keV}$ with
  a$_{XMM}$=10.3$^{+1.1}_{-1.2}\times$10$^7$ and norm$_{Fe
    K}$=b$\times$Flux$_{0.5-10 keV}$ with
  b$_{XMM}$=10$^{+5}_{-4}\times$10$^7$ ph. ergs$^{-1}$ during the
  \xmm\ observation.  While the relations are:
  a=23.9$\pm1.5\times$10$^7$ and b=20$\pm5\times$10$^7$
  ph. ergs$^{-1}$ during the \suzaku\ one.  The variations of the Fe L
  and K lines are thus consistent with being the same within each
  observation. This suggests that a unique reflection component can
  reproduce the two features at all flux states during the two
  observations. However, during the \xmm\ observation the intensities
  of the lines (filled blue and open light blue stars) are higher than
  during the \suzaku\ observation (red filled and orange open squares)
  at the same flux level. This suggests that the reflection component
  (that is tracked by the emission lines) is correlated with the
  continuum within the single observations, while for the same
  continuum flux the \xmm\ observation shows a higher amount of
  reflection.  Nevertheless, the same ratio between Fe L and Fe K is
  always maintained. 

\begin{figure} 
 \includegraphics[width=0.46\textwidth,height=0.34\textwidth]{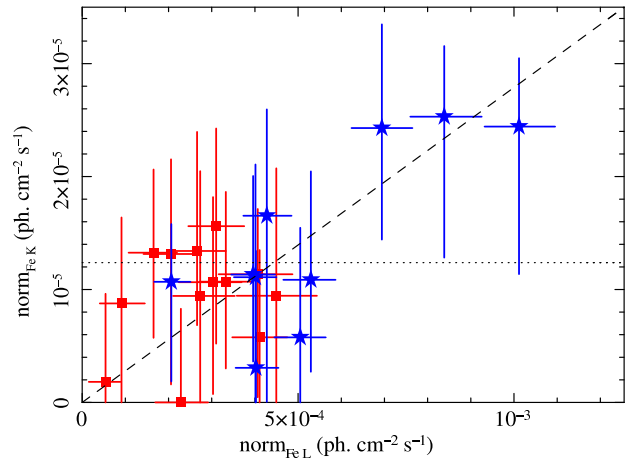}
\caption{Intensity of Fe L vs. Fe K. The intensities of the two lines
  are varying in a correlated way. The variations are consistent with
  a constant ratio between the two features. This suggests that the
  two lines are produced by a common spectral component. The best fit
  relation is: norm$_{Fe K}$=a$\times$norm$_{Fe L}$ with
  a=0.028$\pm0.006$ indicating a ratio of 35$\pm$6, consistent with
  the values expected from disc reflection (Ross \& Fabian 2005).}
\label{FeLFeK}
\end{figure}
Figure \ref{FeLFeK} confirms this trend. In fact although we
  observe that a fit with a constant is acceptable ($\chi^2=18.8$ for
  22 dof), a linear correlation between the intensities of the two
lines provides a better fit than a constant ($\Delta\chi^2$=10.2 for
the same dof; $\chi^2=8.6$ for 22 dof). This indicates that the
two main spectral features at low ($\sim$1.2 keV) and high ($\sim$8
keV) energies are consistent with having a common origin.
% and that their fluctuations are
%connected, most probably because they originate from the same spectral
%component. 
The best fit coefficient of the
linear relation is: norm$_{Fe K}$=a$\times$norm$_{Fe L}$; with
a=0.028$\pm0.006$ (see Fig. 8) indicating a ratio between Fe K and Fe
L of 35$\pm$6, at all flux levels. 
%This value is consistent with the
%one expected from ionised disc reflection (Ross \& Fabian 2005).
%fit 4 (dati blue)
%\chi^2=8.6 for 22 dof
%0.028+-0.006

\subsection{Ionized disc reflection}

We fitted the spectra with a physically self-consistent model 
including a power law and an ionised disc reflection component.
%We decided to study both models, and not only the more
%physically plausible one, in order to better understand the robustness
%of the results.  Thus, each of the ten spectra has been fitted twice
%with a power law with fixed slope and variable normalization plus an ionized
%disc reflection, the spectrum of which is convolved with the same
%{\footnotesize{LAOR}} kernel of the mean spectrum. Clearly the ionized
%reflection had in the former case a low energy cut off at 1 eV while
%in the latter one at 22 eV. In the fit the only free parameter of the
%relativistic kernel is the index q of the inner disc emissivity
%profile ($\epsilon = r^{-q}$). The ionized reflection model has the
%normalization as free parameter (the disc ionisation is fixed to 1000
%and 200 erg cm s$^{-1}$, respectively), while the photon index of the
%illuminating power law in the ionized reflection model is tied to that
%of the power law and therefore fixed to $\Gamma\sim$3 and 3.12,
%respectively.  In both cases the model reproduce very well the data at
%all flux levels with a reduced $\chi^2$ between 0.9 and 1.2.
We assume that the inner and outer disc radius; inclination; and the
elemental abundances do not vary and we fix them to the best fit
values of the mean \xmm\ and \suzaku\ models (consistent with being
the same). We also fix the power law spectral index. Thus, the only
free parameters are the normalisation of the two components, the disc
ionisation and the emissivity index.  This model reproduces well the
spectra at all flux levels.  From the best fit results of the time
resolved spectra we simulated the expected rms spectra. The red line
in Fig. 3 shows that this decomposition reproduces the source spectral
variations during both observations.  Despite some discrepancies, the
general rms spectral shape is well reproduced by this minimal two
component model. Further model complexity is required to accurately
reproduce the rms features and it is beyond the scope of this paper
which seeks a minimal-complexity solution.

\begin{figure}
\includegraphics[width=0.46\textwidth,height=0.34\textwidth]{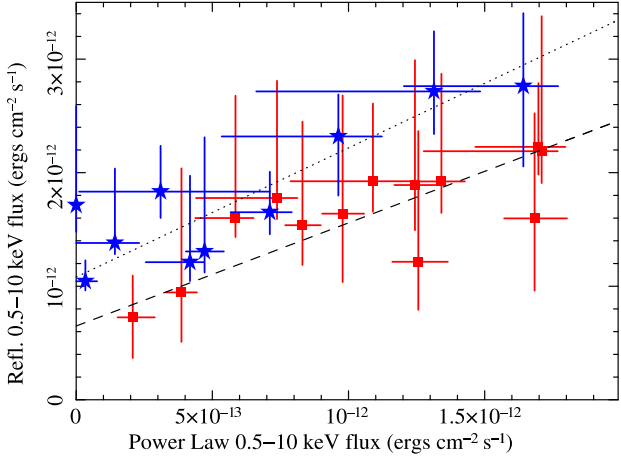}
\caption{Ionised disc reflection versus power law component
  0.5--10~keV fluxes.  Both data sets give a superior fit with a
  linear model than with a constant at 99 \% confidence limit (dotted
  and dashed line for \xmm\ and \suzaku\ data, respectively). In
  particular the best fit relation is consistent with having the same
  amplitude of the variations of the two components. A positive
  intercept of the y-axis is observed extrapolating the best fit
  relation at null power law fluxes. }
\label{PLC RDC}
\end{figure}

Fig. \ref{PLC RDC} shows the 0.5--10~keV flux of the ionised
reflection versus the power law flux in the same band.  During both
observations a fit with a straight line is superior to a constant at a
99 \% confidence level.  The linear Pearson correlation coefficient is
$r=0.75$ and $r=0.87$ for the \xmm\ and \suzaku\ points,
respectively. In both cases, the spectra appear to be
reflection--dominated, most remarkably at low flux levels.

The correlation between the two spectral components
%is consistent with equal variability amplitude 
%in the power law and reflection component.
%Nevertheless, both fits 
requires a positive intercept in the y-axis.  Physically this
indicates the presence of an ionised disc reflection component even
when the power law flux is null.  In particular the main difference
between the two observations is that in 2002 the amount of this 
  residual reflection component was higher
($\sim$1.1$\times$10$^{-12}$ ergs cm$^{-2}$ s$^{-1}$ instead of
$\sim$6$\times$10$^{-13}$ ergs cm$^{-2}$ s$^{-1}$).  The presence of
this difference in the intensity of the residual reflection
component explains also the difference in the shapes of the rms
spectra (see Fig. 3).

Reflection--dominated spectra have been observed in many AGN and in
particular in NLS1 accreting at a high rate (Zoghbi et al. 2008;
Schartel et al. 2007; Fabian et al. 2004; Grupe et al. 2008).  This
can happen for example when disc instabilities break the accretion
disc in rings of dense material and the primary source, inside these,
is thus hidden from view leaving only a reflection component (Fabian
et al. 2002).  A reflection dominated spectrum is also expected when
the nuclear source is situated only a few gravitational radii from the
black hole.  In this conditions light bending effects are strong,
generating (as seen at infinity) a strong reflection and a dim primary
source (Miniutti et al. 2003; Miniutti \& Fabian 2004). In the most
extreme cases (regime I in Miniutti \& Fabian 2004) corresponding to
strongly reflection--dominated spectra (which is the relevant case
here) the light bending model implies steep emissivity profiles and
correlated variability between the reflection and primary (power law)
components. Although the latter is a prediction of the model for
reflection--dominated states (as opposed to less reflection--dominated
cases in which the model predicts a more contant reflection),
correlated variability between the primary and reprocessed component
is what is generally expected in any reprocessing model and cannot be
associated unequivocally with the specific variability model of Miniutti
\& Fabian (2004).

\begin{figure}
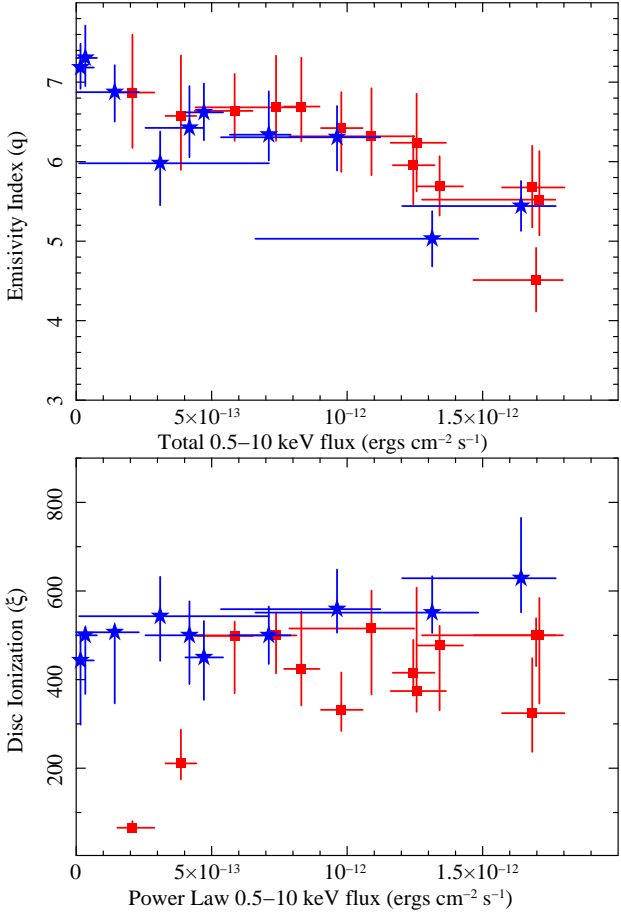

\includegraphics[width=0.34\textwidth,height=0.46\textwidth,angle=-90]{Flux0510_Index_xis_err_mod.ps}
\includegraphics[width=0.34\textwidth,height=0.46\textwidth,angle=-90]{Flux0510_Xi_xis_err200_mod.ps}
\caption{{\it Upper panel } The emissivity index of the blurring
  profile is plotted vs. the  0.5--10 keV power law flux. {\it Lower
    panel} Disc ionisation parameter vs. power law flux.}
\label{emis}
\end{figure}

The upper panel of Fig.~\ref{emis} shows the disc
reflection emissivity index $q$ as a function of the 0.5--10~keV power
law flux.  The emissivity index appears to be weakly anti-correlated
with flux (linear Pearson correlation coefficient of $r=-0.85$), which
is in line with the light bending model predictions in which low flux
states are automatically associated with more centrally concentrated
disc illumination profiles. However, even for the most extreme light
bending effects, emissivity indices as steep as 6-7 are difficult to
produce and hence somewhat suspicious. Here we point out only one
possible reason: the relativistic blurring model we use is based on
the Laor (1991) code and assumes limb darkening, while numerical
simulations would in fact suggest that mild limb brightening is more
appropriate for disc reflection (Svoboda et al. 2009). In fact,
Svoboda et al. (2009) have shown that, by applying a limb brightning
model to the best-studied broad line case of MCG-6-30-15, the inner
emissivity index can be reduced from $\sim$5.3 to $\sim$3.7 for similar
inclination and black hole spin.  The emissivity index may thus be
generally overestimated, especially at relatively high inclination
angles such as the one inferred in the present case.

\begin{figure}
\includegraphics[width=0.34\textwidth,height=0.46\textwidth,angle=-90]{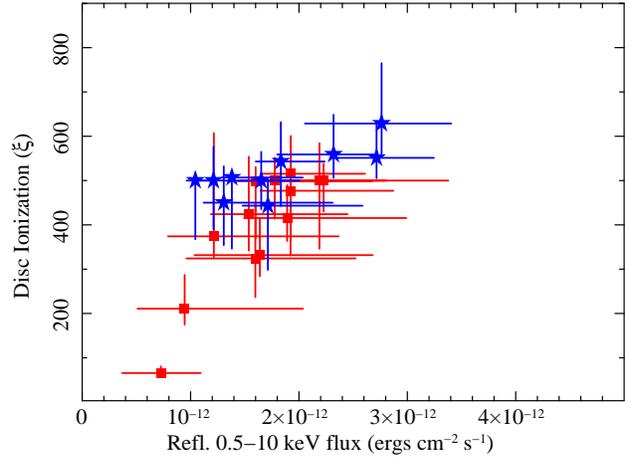}
\caption{The disc ionization as a function of the reflection component
  flux. The observed correlation provides a sanity check for our
  reflection interpretation because both quantities should respond to
  the same physical quantity, i.e. the X--ray flux irradiating the
  accretion disc.}
\label{xirefl}
\end{figure}

The lower panel of Fig.~12 shows the disc ionisation parameter as a
function of the direct (power law) flux. In this case the correlation
is very poor ($r=0.20$) meaning that the variations of the disc
ionisation parameter do not closely follow the variations of the
source flux (a similar behaviour is present in other sources,
i.e. MCG-6-30-15; Ballantyne et al. 2003). We point out that, if light
bending really plays a role in shaping the general spectral and
variability properties of the source, the observed flux cannot be
directly connected to the flux illuminating the accretion disc (which
is the one to which the ionization should be associated with). This is
because, due to light bending, the disc sees a different flux than the
observer at infinity. On the other hand, both the disc ionization
state and the reflection intensity should be directly correlated with
the flux irradiating the accretion disc. Hence a correlation between
reflection intensity and ionization state is generally expected,
especially if the bulk of the reflection component comes from a
limited region (e.g. the innermost accretion disc). Such a trend is
indeed observed in our data (see Fig.~\ref{xirefl}), and represents a
sanity check for our reflection interpretation ($r=0.65$).

IRAS13224-3809 is not the only source where the nuclear emission is
thought to originate from within the regions where strong gravity
effects are not negligible (Fabian \& Miniutti 2005).  Long
uninterrupted exposures of such AGN have demonstrated that the
relation between direct emission and the disc reflection component is
complex. The cases of MCG-6-30-15 and NGC4051 show that the disc
reflection component is approximately constant at medium/high fluxes
and correlated with the power law at low fluxes where the spectrum
becomes more reflection--dominated (Ponti et al. 2006; Larsson et
al. 2007).  This is the expected behaviour when the intrinsic
luminosity variations of the X-ray source are negligible and most of
the flux variability is due to general relativistic effects.  A longer
observation of IRAS13224-3809 is required to fully address this issue.

\subsubsection{Fe L / K energies and ratio}

The ratio between Fe L and Fe K depends on many physical parameters,
primarily on the power law index of the illuminating source and on the 
physical properties of the reflecting material.
Kallman (1995) simulated the expected ratio assuming a slab of optically 
thin gas with Solar abundances photo-ionised by a X-ray power law source 
with an energy index of -1.5. In such conditions the Fe L/ FeK ratio is 
expected to be about 3 for an ionisation parameter $log(\xi)\sim2-3$. 
To measure the ratio in the case of an optically thick accretion disc with 
a high Fe abundance, we computed extensive simulations of accretion 
disc in the same scenario used to compute the ionised reflection 
model (Ross \& Fabian 2005). In particular the primary X-ray source 
illuminating the accretion disc has a power law spectral index 
of $\Gamma=2.4$, the disc has a Fe abundance 5 times higher than solar 
and ionisation $\xi=500$ erg cm s$^{-1}$. 
The ratio between the intensities of Fe L and K has been calculated considering 
the Fe L region extending from 0.71 keV to 1.08 keV, and the Fe K region 
to extend from 6.31 keV to 7.07 keV (where the smeared Fe K-edge begins). 
The continuum has been fitted with a simple straight-line continuum in these 
two regions and then added up the line photons (how much the emergent 
spectrum exceeded the continuum fit) for each of these regions. 
We found a Fe L to Fe K photon ratio of 44.6 for this model. 
The reason that the L line dominates so strongly in this model is
that Fe XVII dominates over most of the outermost Thomson depth, 
while the K line is suppressed by Auger destruction during resonance 
trapping. We also observe that, varying the ionisation parameter 
between 200 and 600 erg cm s$^{-1}$, the FeL/K ratio has 
values of a few tens. 
These computations depend on the assumption of the straight line 
continuum and thus do not provide the precise FeL/K ratio. Nevertheless,
they should provide reasonable values to compare with the phenomenological 
broad lines model where the continuum is also a straight line.
We, in fact, observe that these values are roughly in agreement with the ones 
obtained from the fit with broad lines (see $\S$6).

\section{Frequency dependent lags}

\begin{figure}
\includegraphics[width=0.46\textwidth,height=0.34\textwidth]{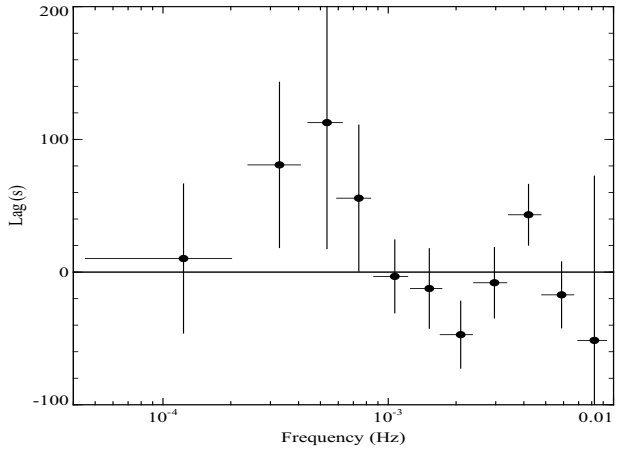}
\caption{Frequency-dependent lags between the 0.3--1 and 1--4 keV bands. 
Negative lags indicate the harder flux (dominated by the power law continuum) 
changes before the softer flux (dominated by reflection, i.e. the iron L line). 
The timing properties of IRAS13224-3809 appear very similar to the ones 
of 1H0707-495 (Fabian et al. 2009), nevertheless the shortness of the observation
hamper a significant detection of the lags.}
\label{lag}
\end{figure}

IRAS13224-3809 is well known to present a high degree of complex
variability (see Fig. \ref{fig:lcs} and \ref{fig:rms}). Gallo et
al. (2004) studied the timing properties investigating the
correlations and showing a complex pattern of lags and leads between
the 0.3-0.8 and 3-10 keV band (see Fig. 4 of Gallo et al. 2004).  In
combination with the spectral information, we performed a time-lag
study (of the uninterrupted \xmm\ data) separating the energy regions
dominated by the power law and ionised reflection emission.  Figure
\ref{lag} shows the frequency dependent lags between the
reflection-dominated 0.3-1 keV and the power law-dominated 1-4 keV
bands (see Fig. 8). Negative lags indicate that the hard (power law
dominated) components leads the reflection emission. The shape of the
frequency dependent lags closely resembles the one of 1H0707-495,
where for the first time a reverberation lag of 30 s between the
continuum and the energy band of the Fe L reflection line is detected
(Fabian et al. 2009).  In the present case a significant detection of
the lag is prevented by the short exposure (60 ks, about ten times
lower than 1H0707-495). A long uninterrupted observation of
IRAS13224-3809 will clarify this issue and serve to scrutinise the
reflection dominated interpretation proposed here.

\section{Discussion}

We have studied the X--ray spectral variability of \iras\ analyzing
the \xmm\ and \suzaku\ data. The source shows very similar behaviour 
during the two observations taken 5 years apart.
In particular it exhibits: i) a strong soft excess with a sharp feature around 
1.2~keV; ii) an impressive drop in flux at $\sim$8.2~keV (the drop has 
not shifted in energy and no narrow component of the Fe K line is detected).

These features may be reproduced by either absorption or ionized
reflection. In the absorption interpretation the two features imply at
least two distinct absorption components, while in the reflection
interpretation a single reflection component can explain all the major
spectral features. Moreover, the reflection model naturally
accounts also for the soft excess, while additional soft
(e.g. blackbody) components must be included in the
absorption--dominated models we have considered. In fact when the
spectrum is fitted with two broad lines, the energies of the lines are
consistent with coming from Fe~L and Fe~K, the shapes are consistent
with being produced by the same broadening profile and the ratio
between the intensity broadly consistent with expectations from a
  dominant Fe XVII contribution. This result is very similar to what
has been recently reported for a similar object 1H0707-495 (Fabian et
al. 2009). In that case the reflection interpretation is strengthened
by the detection of a time lag between the soft band (0.3-1 keV,
dominated by the Fe L emission) and the power law, hampered here by
the shortness of the \xmm\ observation.  Longer uninterrupted
observations are required to study the timing characteristics of
\iras\ in detail.

We stress, however, that ionised disc reflection not only gives a
reasonable fit to the mean spectrum, but also reproduces the source
spectral variability. In particular the variability can be described
by a variable power law with constant spectral shape and a reflection
component following the power law variations, with residual reflection
dominating the spectrum at low fluxes. During the \xmm\ observation
this residual ionised disc reflection component is more
prominent even if the source flux is similar.

The reflection component is affected by relativistic effects, strongly
suggesting that it originates in the innermost regions of the
accretion disc around a rapidly rotating Kerr black hole. The best fit
inner radius of the accretion disc is of the order of 1.2 r$_g$ with a
90\% upper limit of $\sim$2~r$_g$, implying an almost maximal black
hole spin $a>0.94$. The disc illumination radial profile is consistent
with a steep power law (q$\sim$5--7).  A weak anti-correlation between
the illumination profile and the source direct flux is observed. This
indicates that at lower fluxes the primary emission is likely closer
to the back hole. During the low flux states most of the reflection is
produced within a few gravitational radii from the black hole. In such
conditions strong relativistic effects must be at work.

Despite being an illustrative toy model, which most likely represents
the zeroth-order approximation of the full relativistic cases, the
light bending model proposed by Miniutti \& Fabian (2004) seems to
capture the main characteristics of \iras. In particular the model
predicts the existence of a regime in which the spectrum is strongly
reflection-dominated, as we observe here. Within this regime, the
reflection and continuum components are expected to correlate well,
while the disc emissivity is anti--correlated with the direct flux
(the power law flux in this case). Although we observe both trends, we
point out that correlated variability between the primary and
reprocessed component is a general prediction of any reflection
model. However, the light bending model associates this behaviour only
with reflection--dominated states, while in more standard cases, the
reprocessed component is predicted to vary less than the
primary. Future X--ray observation catching the source in less
reflection--dominated states may be important to test the model
predictions.

Both the strong soft excess and the extreme Fe K lines, when fitted 
with a disc ionised reflection component, require an extremely high level 
of reflection, near to have a reflection dominated spectrum (see Fig. 8).
We measure a ratio between the ionised reflection to the power law flux of 
1.7 and 0.75 over the 0.001-1000 keV band (the entire band over which 
the reflection model is computed), while the ratio is 3.7 and 1.6 in the 
observed 0.5-10 keV band for the \xmm\ and \suzaku\ observations, respectively.
We note that the high energy part of the reflected spectrum is 
less affected by ionisation, thus we estimated the reflection fraction 
comparing the extrapolated best fit ionised reflection to power law 
flux ratio in the 20-60 keV band (that results to be about 4.9 and 2.1) 
with the one expected by neutral reflection (using a {\sc pexrav} 
component). 
A reflection fraction of 7 and 4 are thus estimated in this way.
\begin{figure}
\includegraphics[width=0.34\textwidth,height=0.46\textwidth,angle=-90]{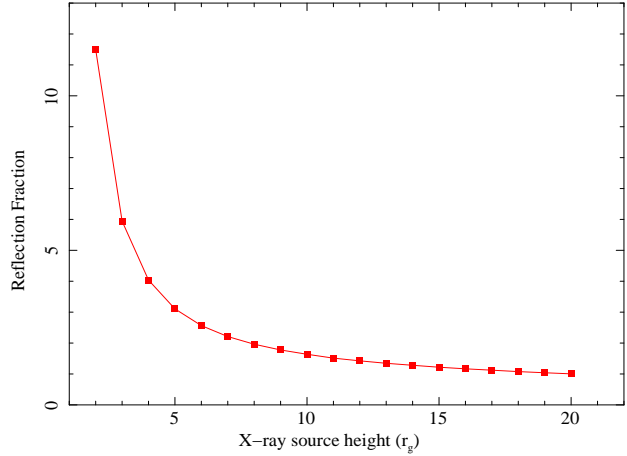}
\caption{Expected reflection fraction as a function of the source
    height above the black hole assuming a on-axis X--ray primary
    source including light bending (Fabian \& Miniutti 2004). The
    original light bending model only considers continuum and Fe line
    production, not the reflection continuum. Thus the reflection
    fraction shown here has been computed by rescaling the calculated
    Fe line equivalent width (EW) with respect to the continuum only
    using the fact that $EW\propto R$, and by assuming that at large
    hights, where light bending is negligible $R\equiv 1$, as
    standard.}
\label{R}
\end{figure}
Are such high reflection fractions consistent with the picture
described in the light bending model?  As shown in Figure 15 the
observed values are rather extreme, however, they are expected when
the primary X-ray source is less than 3-4 r$_g$ above the central
black hole (regime I, see Miniutti \& Fabian 2004).  We estimate that,
in this condition, only about 10--20 \% of the primary X-ray source is
detected at infinity in the form of the power law emission.  Thus, the
measured 2-10 keV power law luminosity of $2-4\times10^{42}$ erg
s$^{-1}$ would correspond to an intrinsic luminosity (in the frame of
the source) of the order of a few times $10^{43}$ erg s$^{-1}$.  This
value is not too large if compared to the bolometric luminosity
(L$_B\sim6\times10^{45}$ erg s$^{-1}$), even assuming a rather extreme
bolometric correction (Elvis et al. 1995; Vasudevan et al. 2007;
2009).  Also in agreement with such a scenario, we observe that the
reflection fraction and the broadness of the line increase lowering
the flux, passing from the \suzaku\ to the \xmm\ observations.

%On the other hand,
%the reflector ionisation state in this regime should be correlated
%with flux at the very low flux levels reaching a plateau and then a
%decay at higher fluxes. All these features are observed in \iras,
%which makes it most likely that X-rays are emitted from within $\sim$4
%r$_g$ from the central rapidly rotating Kerr black hole.

We also observe a poor correlation between disc ionization and
observed power law flux. In the framework of the light bending model,
this is not highly surprising since the disc sees a different
continuum than the observer precisely because of light
bending. However, the disc ionization and the reflection intensity
track each other reasonably well, which is consistent with our
reflection interpretation since they both respond to the flux
irradiating the disc.

The strong ionised reflection interpretation requires a very high 
value of the disc illumination parameter, but also of the
iron abundance ($\sim$5 times solar). 
Nevertheless it is important to stress that an even higher iron 
abundance is required in the partial covering interpretation (Boller et
al. 2003). Moreover supersolar metal abundances of the 
environment around supermassive black holes seem to be the 
rule, more than the exception. For example, the quasar environments 
are metal-rich with typically gas abundances several times Solar at
all redshifts (Hamann et al. 2007). In fact, several methods based on 
optical-UV broad and narrow emission and absorption lines agree 
in observing supersolar metal abundances in large samples of quasars. 
The typical metallicity is observed to be roughly 2-4 times Solar 
(Dietrich et al. 2003a,b; Nagao et al. 2006; Groves et al. 2006; D'Odorica 
et al. 2004), in agreement with the value observed in IRAS13224-3809. 
Our Galactic Centre, even if it is not hosting, at the moment (Ponti et al. 
2010), an accreting black hole, contains large molecular clouds with Fe 
abundances higher than Solar. The Fe K line emitted by Sgr B2, 
for example, suggest a Fe over abundances of the order of 2, maybe 
more (Revnivtsev et al. 2004; Terrier et al. 2010).
Abundances studies in samples of NLS1 suggest that the metals are 
about 2-3 times higher than Solar in these sources (ses Nagao et al. 2002). 
In particular, a five times Solar metallicity has been measured in other 
nearby Narrow Line Seyfert galaxies (i.e. Mrk 1044, Fields et al. 2005). 
The optical/UV spectrum of IRAS13224-3809 presents some of the strongest 
Fe emission lines of the NLS1 class.
The Fe overabundance requires a high rate of explosion of type Ia supernovae
that produce much more iron (more than an order of magnitude) than
type II ones (Nomoto et al. 1997a,b). This may happen in the presence 
of a nuclear star cluster that formed white dwarfs that, through
close interactions, become part of close binaries, enriching the 
environment with Iron through many SN Ia 
explosions (Shara \& Hurley 2002; Fabian et al. 2009).

\section{Conclusions}

\iras\ is a remarkable source, with one of the strongest soft excesses
observed, and two prominent features at $\sim$1.2 and $\sim$8.2 keV.

\begin{itemize}
\item{} These features can be reproduced by 2 broad lines, in
  particular: i) the best fit energies of the lines are consistent
  with emission from Fe~L and Fe~K; ii) the same relativistic profile
  can explain the shape of both lines; iii) the ratio between the
  intensities of the lines is broadly in agreement with expectations.
\item{} A single ionized disc reflection model (produced in the inner accretion
disc) can reproduce the main features 
of the spectrum, implying a iron abundance of $\sim$5 times solar.
%\item{} Nevertheless, the fit with an ionized disc reflection component is 
%unable to reproduce all the details of the mean spectrum leaving some 
%wiggles, in particular in the soft band. 
%A better fit has been obtained by Boller et al. 2003 who fitted the spectrum 
%with many partial covering components. However they used five different 
%and separate components, many of which operate on separate parts of the 
%spectrum, whereas the reflection model has only two components both of 
%which are broad band. In the reflection interpretation, the assumption of 
%Solar abundances for all the elements apart from iron and a distribution 
%in the ionization of the disc may be the cause of the remaining wiggles.
\item{} The main spectral variations can be reproduced by a steep
  power law varying in normalisation and a reflection component from
  the inner accretion disc varying in a correlated way, but with a
  residual component, dominating at low fluxes.
\item{} The spectral features may, in an alternative interpretation,
  be associated with absorption components. Nevertheless in this
  scenario the two main features require at least two physically
  different absorbing components in highly relativistic outflow and
  without a clear physical link. A high iron abundance is also
  required. Moreover, absorption-dominated models over--predict Fe M
  absorption in the soft X--rays, and they still require a likley
    unphysical additional blackbody component to describe the soft
    excess, casting further doubts on the overall interpretation.
  
\end{itemize}

\section*{Acknowledgments}

The work reported here is based on observations obtained with
XMM-Newton, an ESA science mission with instruments and contributions
directly funded by ESA Member States and NASA, and with Suzaku, 
a collaborative mission between the space agencies of Japan (JAXA) 
and the USA (NASA). GP thanks Fabio Mattana, Michael Mayer, 
Massimo Cappi and Mauro Dadina for many useful comments, 
suggestions and technical help. GP thanks ANR for support (ANR-06-JCJC-0047). 
ACF thanks the Royal Society and RRR thanks the college of the Holy 
Cross for support. GM thanks the Spanish Ministerio de Ciencia e 
Innovaci\'on and CSIC for support through a Ram\'on y Cajal contract 
and the Spanish Ministerio de Ciencia e Innovaci\'on for partial support 
through project ESP2006--13608--C02--01. WNB thanks NASA for 
support through LTSA grant NAG5-13035.

\end{document}